\documentclass[12pt]{article}

\usepackage{graphicx,latexsym,amssymb,amsmath}

\newcommand{\edo}{\end{document}}
\newcommand{\comment}[1]{}

\newcommand{\reftheorem}[1]{Theorem~\ref{#1}}
\newcommand{\reflemma}[1]{Lemma~\ref{#1}}
\newcommand{\refprop}[1]{Proposition~\ref{#1}}

\newcommand{\R}{{\mathbb R}}

\newcommand{\N}{{\mathbb N}}

\newcommand{\Rpn}{\R_+^n}
\newcommand{\Rp}{\R_+}
\newcommand{\sign}{\mbox{sign}\,}
\newcommand{\barx}{\bar{x}}
\newcommand{\bary}{\bar{y}}
\newcommand{\barz}{\bar{z}}

\newtheorem{theorem}{Theorem}
\newtheorem{itlemma}{Lemma}[section] 
\newtheorem{itproposition}[itlemma]{Proposition}
\newtheorem{itcorollary}[itlemma]{Corollary}
\newtheorem{itremark}[itlemma]{Remark}
\newtheorem{itdefinition}[itlemma]{Definition}
\newtheorem{itexample}[itlemma]{Example}

\newenvironment{lemma}{\begin{itlemma}\rm}{\end{itlemma}} 
\newenvironment{remark}{\begin{itremark}\rm}{\end{itremark}} 
\newenvironment{corollary}{\begin{itcorollary}\rm}{\end{itcorollary}}
\newenvironment{proposition}{\begin{itproposition}\rm}{\end{itproposition}}
\newenvironment{definition}{\begin{itdefinition}\rm}{\end{itdefinition}}
\newenvironment{example}{\begin{itexample}\rm}{\end{itexample}}

\newcommand{\be}[1]{\begin{equation}\label{#1}}
\newcommand{\ee}{\end{equation}}
\newcommand{\bl}[1]{\begin{lemma}\label{#1}}
\newcommand{\ble}[1]{\begin{lemmaex}\label{#1}}
\newcommand{\br}[1]{\begin{remark}\label{#1}}
\newcommand{\bt}[1]{\begin{theorem}\label{#1}}
\newcommand{\bd}[1]{\begin{definition}\label{#1}}
\newcommand{\bp}[1]{\begin{proposition}\label{#1}}
\newcommand{\bc}[1]{\begin{corollary}\label{#1}}
\newcommand{\bfact}[1]{\begin{fact}\label{#1}}
\newcommand{\ber}[1]{\begin{exercise}\label{#1}}
\newcommand{\bex}[1]{\begin{example}\label{#1}}
\newcommand{\bem}[1]{\begin{example}\label{#1}}  
\newcommand{\ec}{\mybox\end{corollary}}
\newcommand{\efact}{\mybox\end{fact}}
\newcommand{\eer}{\mybox\end{exercise}}
\newcommand{\eex}{\mybox\end{example}}
\newcommand{\eem}{\mybox\end{example}}
\newcommand{\el}{\mybox\end{lemma}}
\newcommand{\ele}{\mybox\end{lemmaex}}
\newcommand{\er}{\mybox\end{remark}}
\newcommand{\et}{\qed\end{theorem}}
\newcommand{\ed}{\mybox\end{definition}}
\newcommand{\ep}{\mybox\end{proposition}}
\newcommand{\epr}{\end{proof}}
\newcommand{\bpr}{\begin{proof}}

\newcommand{\ecs}{\end{corollary}}
\newcommand{\eers}{\end{exercise}}
\newcommand{\eexs}{\end{example}}
\newcommand{\eems}{\end{example}}
\newcommand{\els}{\end{lemma}}
\newcommand{\eles}{\end{lemmaex}}
\newcommand{\ers}{\end{remark}}
\newcommand{\ets}{\end{theorem}}
\newcommand{\eds}{\end{definition}}
\newcommand{\eps}{\end{proposition}}
\newcommand{\halmos}{\rule{1ex}{1.4ex}}
\newcommand{\qed}{\hfill \halmos} 
\newcommand{\mybox}{\hfill $\Box$} 
\newcommand{\beq}{\begin{eqnarray}}
\newcommand{\eeq}{\end{eqnarray}}
\newcommand{\beqn}{\begin{eqnarray*}}
\newcommand{\eeqn}{\end{eqnarray*}}
\newcommand{\bi}{\begin{itemize}}
\newcommand{\ei}{\end{itemize}}
\newcommand{\ben}{\begin{enumerate}}
\newcommand{\een}{\end{enumerate}}

\newenvironment{proof}{\noindent {\em Proof}.\ }{\hspace*{\fill}$\halmos$\medskip}

\title{A Petri net approach to the study of persistence\\
 in chemical reaction networks}
\author{David Angeli\footnote{Email: angeli@dsi.unifi.it}\\ Dip. di Sistemi e Informatica, University of Firenze \\ \newline \\
Patrick De Leenheer\footnote{Email: deleenhe@math.ufl.edu.
Supported in part by NSF Grant NSF DMS-0614651}\\ Dep. of Mathematics, University of Florida,
Gainesville, FL\\
\newline \\ Eduardo D. Sontag\footnote{Corresponding author.
Email: sontag@control.rutgers.edu. %
Supported in part by NSF Grant NSF DMS-0504557}\\Dep. of Mathematics, Rutgers
University, Piscataway, NJ}

\begin{document}
\maketitle

\begin{abstract}

  Persistency is the property, for differential equations in $\R^n$, that
  solutions starting in the positive orthant do not approach the boundary.
  For chemical reactions and population models, this translates into the
  non-extinction property: provided that every species is present at the start
  of the reaction, no species will tend to be eliminated in the course of the
  reaction.  This paper provides checkable conditions for persistence of
  chemical species in reaction networks, using concepts and tools from Petri
  net theory, and verifies these conditions on various systems which arise in
  the modeling of cell signaling pathways.

\end{abstract}

\newpage

\section{Introduction}

One of the main goals of molecular systems biology is the understanding of
cell behavior and function at the level of chemical interactions, and, in
particular, the characterization of qualitative features of dynamical
behavior (convergence to steady states, periodic orbits, chaos, etc).
A central question, thus, is that of understanding the long-time behavior of
solutions.  In mathematical terms, and using standard chemical kinetics
modeling, this problem may be translated into the study of the set of possible
limit points (the \emph{$\omega$-limit set}) of the solutions of a system of
ordinary differential equations.

\subsubsection*{Robustness}

A distinguishing feature of this study in the context of cell biology, in
contrast to more established areas of applied mathematics and engineering, is
the very large degree of uncertainty inherent in models of cellular
biochemical networks.  This uncertainty is due to environmental fluctuations,
and variability among different cells of the same type, as well as, from a
mathematical analysis perspective, the difficulty of measuring the relevant
model parameters (kinetic constants, cooperativity indices, and many others)
and thus the impossibility of obtaining a precise model.  Thus, it is
imperative to develop tools that are ``robust'' in the sense of being able to
provide useful conclusions based only upon information regarding the
\emph{qualitative} features of the network, and not the precise values of
parameters or even the forms of reactions.  Of course, this goal is often
not unachievable, since dynamical behavior may be subject to phase
transitions (bifurcation phenomena) which are critically dependent on
parameter values. 

Nevertheless, and surprisingly, research by many, notably by
Clarke~\cite{clarke},
Horn and Jackson~\cite{hornjackson1,hornjackson2},
Feinberg~\cite{feinberg0,feinberg1,feinberg2},
and many others in the context of complex balancing and deficiency theory,
and by Hirsch and Smith~\cite{hal1,hal2} and many others including the present
authors~\cite{our1,our2,our3,our4}
in the context of monotone systems, has resulted in the identification of rich
classes of chemical network structures for which such robust analysis is
indeed possible.
In this paper, we present yet another set of tools for the robust analysis of
dynamical properties of biochemical networks, and apply our approach in
particular to the analysis of persistence in chemical networks modeled by
ordinary differential equations.
Our approach to studying persistence is based on the formalism and basic
concepts of the theory of Petri nets.  Using these techniques, our main
results provide conditions (some necessary, and some sufficient) to test
persistence.  We then apply these conditions to obtain fairly tight
characterizations in non-trivial examples arising from the current molecular
biology literature.

\subsubsection*{Persistency}

\emph{Persistency} is the property that, \emph{if every species is present at
  the start of the reaction, no species will tend to be eliminated
in the course of the reaction}. 
Mathematically, this property can be equivalently expressed as the requirement
that the $\omega$-limit set of any trajectory which starts in the interior of
the positive orthant (all concentrations positive) does not intersect the
boundary of the positive orthant (more precise definitions are given below).
Persistency can be interpreted as non-extinction: if the concentration of a
species would approach zero in the continuous differential equation model, this
means, in practical terms, that it would completely disappear in finite time,
since the true system is discrete and stochastic.
Thus, one of the most basic questions that one may ask about a chemical
reaction is if persistency holds for that network. 
Also from a purely mathematical perspective persistency is very important,
because it may be used in conjunction with other tools in order to guarantee
convergence of solutions to equilibria.
For example, if a strictly decreasing Lyapunov function exists on the interior
of the positive orthant (see
e.g.~\cite{hornjackson1,hornjackson2,feinberg0,feinberg1,feinberg2,chemTAC}
for classes of networks where this can be guaranteed),
persistency allows such a conclusion.

An obvious example of a non-persistent chemical reaction is a simple
irreversible conversion $A\rightarrow B$ of a species $A$ into a species $B$;
in this example, the chemical $A$ empties out, that is, its time-dependent
concentration approaches zero as $t\rightarrow\infty$.
This is obvious, but for complex networks determining persistency, or lack
thereof, is, in general, an extremely difficult mathematical problem.
In fact, the study of persistence is a classical one in the (mathematically)
related field of population biology, where species correspond to individuals
of different types instead of chemical units;
see for example~\cite{gard,waltman1} and much other foundational work by
Waltman. 
(To be precise, what we call ``persistence'' coincides with the usage in the
above references, and is also sometimes called ``strong persistence,'' at
least when all solutions are bounded, a condition that we will assume in most
of our main results, and which is automatically satisfied in most examples.
Also, we note that a stronger notion, ``uniform'' persistence, is used to
describe the situation where all solutions are eventually bounded away from
the boundary, uniformly on initial conditions, see~\cite{waltman2,thieme}.)
Most dynamical systems work on persistence imposes conditions ruling out
phenomena such as heteroclinic cycles on the boundary of the positive orthant,
and requiring that the unstable manifolds of boundary equilibria should
intersect the interior, and more generally studying the chain-recurrence
structure of attractors, see e.g.  \cite{hofbauer}.

\subsubsection*{Petri nets}

\emph{Petri nets}, also called place/transition nets, were introduced by Carl
Adam Petri in 1962~\cite{ca_petri}, and they constitute a popular mathematical
and graphical modeling tool used for concurrent systems
modeling~\cite{peterson,zhou}. 
Our modeling of chemical reaction networks using Petri net formalism is not in
itself a new idea: there have been many works, at least
since~\cite{reddy},which have dealt with biochemical applications of Petri
nets, in particular in the context of metabolic pathways, see
e.g.~\cite{petri1,petri2,petri3,petri5,petri6}, and especially the
excellent exposition~\cite{schuster}.
However, there does not appear to have been any previous work using Petri nets
for a nontrivial study of dynamics.  In this paper, although we do not use any
results from Petri net theory, we employ several concepts (siphons, locking
sets, etc.), borrowed from that formalism and introduced as needed, in order
to formulate new, powerful, and verifiable conditions for persistence and
related dynamical properties.

\subsubsection*{Application to a common motif in systems biology}

\begin{figure}[ht]
\centerline{%
\includegraphics[scale=0.3]{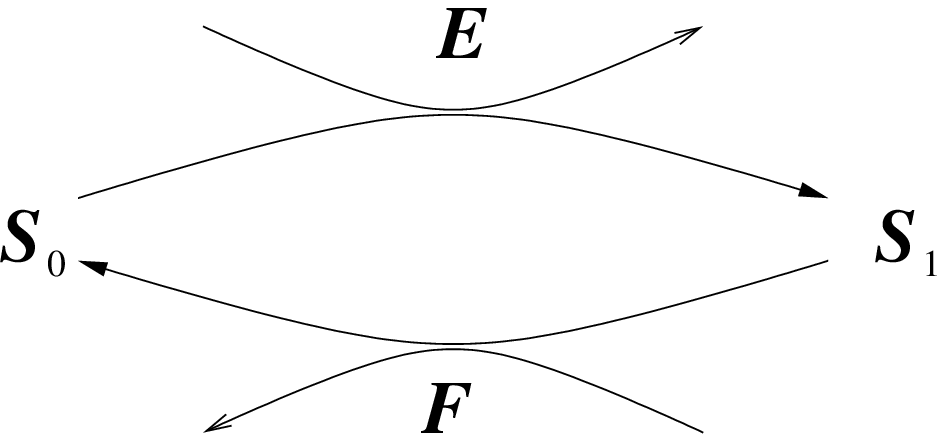}
\hskip3cm
\includegraphics[scale=0.3]{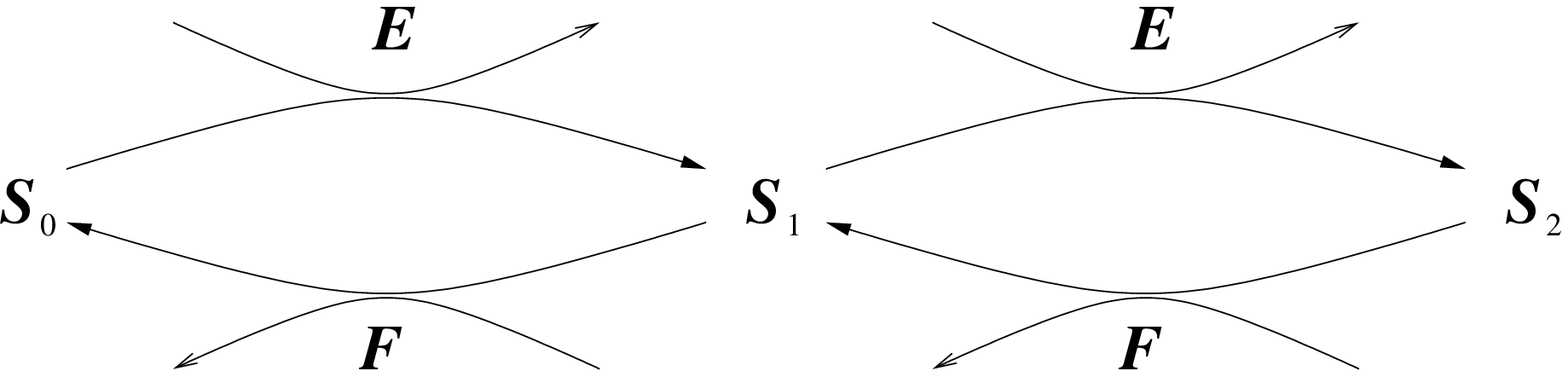}%
}
\caption{\em (a) One-step \hskip3cm and (b) two-step transformations}
\label{futile_cycle}
\end{figure}

In molecular systems biology research, certain ``motifs'' or subsystems appear
repeatedly, and have been the subject of much recent research.  One of the
most common ones is that in which a substrate $S_0$ is ultimately converted
into a product $P$, in an ``activation'' reaction triggered or facilitated by
an enzyme $E$, and, conversely, $P$ is transformed back (or ``deactivated'')
into the original $S_0$, helped on by the action of a second enzyme $F$.  This
type of reaction is sometimes called a ``futile cycle'' and it takes place in
signaling transduction cascades, bacterial two-component systems, and a
plethora of other processes.  The transformations of $S_0$ into $P$ and
vice versa can take many forms, depending on how many elementary steps
(typically phosphorylations, methylations, or additions of other elementary
chemical groups) are involved, and in what order they take place.
Figure~\ref{futile_cycle} shows two examples, (a) one in which a single step
takes place changing $S_0$ into $P=S_1$, and (b) one in which two sequential
steps are needed to transform $S_0$ into $P=S_2$, with an intermediate
transformation into a substance $S_1$.
A chemical reaction model for such a set of transformations incorporates
intermediate species, compounds corresponding to the binding of the enzyme and
substrate.  (In ``quasi-steady state'' approximations, a singular
perturbation approach is used in order to eliminate the intermediates.
These approximations are much easier to study, see e.g.~\cite{our1}.)
Thus, one model for (a) would be through the following reaction network:
\begin{equation}
\label{mapk1}
\begin{array}{ccccccccc}
E + S_0 &\leftrightarrow& ES_0   & \rightarrow   & E + S_1 \\
F + S_1 & \leftrightarrow& F S_1 & \rightarrow   &F + S_0 
\end{array}
\end{equation}
(double arrows indicate reversible reactions)
and a model for (b) would be:
\begin{equation}
\label{mapk}
\begin{array}{ccccccccc}
E + S_0 &\leftrightarrow& ES_0 & \rightarrow & E + S_1 &
\leftrightarrow & ES_1 & \rightarrow &E + S_2 \\
F + S_2 & \leftrightarrow& F S_2 & \rightarrow & F + S_1 & \leftrightarrow &
F S_1 & \rightarrow &F + S_0
\end{array}
\end{equation}
where ``$ES_0$'' represents the complex consisting of $E$ bound to $S_0$ and
so forth.

As a concrete example, case (b) may represent a reaction in which the enzyme
$E$ reversibly adds a phosphate group to a certain specific amino acid in the
protein $S_0$, 
resulting in a single-phosphorylated form $S_1$; in turn, $E$ can then bind to
$S_1$ so as to produce a double-phosphorylated form $S_2$, when a second amino
acid site is phosphorylated.  A different enzyme reverses the process.
(Variants in which the individual phosphorylations can occur in different
orders are also possible; we discuss several models below.)
This is, in fact, one of the mechanisms believed to underlie signaling by MAPK
cascades.  {\em Mitogen-activated protein kinase (MAPK) cascades} constitute a
motif that is ubiquitous in signal transduction processes
\cite{ferrell,lauffenburger,widman} in eukaryotes from yeast to humans, and
represents a critical component of pathways involved in cell apoptosis,
differentiation, proliferation, and other processes.  These pathways involve
chains of reactions, activated by extracellular stimuli such as growth factors
or hormones, and resulting in gene expression or other cellular responses.
In MAPK cascades, several steps as in (b) are arranged in a cascade, with the
``active'' form $S_2$ serving as an enzyme for the next stage.

Single-step reactions as in (a) can be shown to have the property that all
solutions starting in the interior of the positive orthant globally converge
to a unique (subject to stoichiometry constraints) steady state,
see~\cite{trans_invariance_include_conf}, and, in fact, can be modeled by
monotone systems after elimination of the variables $E$ and $F$,
cf.~\cite{cdc06}.  The study of (b) is much harder, as multiple equilibria can
appear, see e.g.~\cite{kholodenko,conradi}.  We will show how our results can
be applied to test consistency of this model, as well as several variants.

\subsubsection*{Organization of paper}

The remainder of paper is organized as follows.
Section~\ref{Chemical Networks} sets up the basic terminology and definitions
regarding chemical networks, as well as the notion of persistence,
Section~\ref{Petri Nets} shows how to associate a Petri net to a chemical
network, 
Sections~\ref{Necessary conditions} and~\ref{Sufficient Conditions} provide,
respectively, necessary and sufficient conditions for general chemical
networks,
In Section~\ref{Applications}, we show how our results apply to the enzymatic
mechanisms described above.

\section{Chemical Networks}
\label{Chemical Networks}

A \emph{chemical reaction network} (``CRN'', for short) is a set of chemical
reactions $\mathcal{R}_i$, where the index $i$ takes values in
$\mathcal{R}:=\{1,2, \ldots, n_r \}$.
We next define precisely what one means by reactions, and the differential
equation associated to a CRN, using the formalism from chemical networks
theory.

Let us consider a set of chemical species $S_j$,
$j \in \{1,2, \ldots n_s \}:=\mathcal{S}$ which are the
compounds taking part in the reactions.
Chemical reactions are denoted as follows: 
\begin{equation}
\label{list}
 \mathcal{R}_i: \quad \sum_{j \in \mathcal{S} } \alpha_{ij} S_j \rightarrow \sum_{j \in \mathcal{S} } \beta_{ij} S_j 
\end{equation}
where the $\alpha_{ij}$ and $\beta_{ij}$ are nonnegative integers called the 
\emph{stoichiometry coefficients}.
The compounds on the left-hand side are usually referred to as the 
\emph{reactants}, and the ones on the right-hand side are called the
\emph{products}, of the reaction. 
Informally speaking, the forward arrow means that the transformation of
reactants into products only happens  in the direction of the arrow. If also
the converse transformation occurs, then, the reaction is 
reversible and we need to also list its inverse in the chemical reaction
network as a separate reaction.

It is convenient to arrange the stoichiometry coefficients into an $n_s\times
n_r$ matrix, called the \emph{stoichiometry matrix} $\Gamma$, 
defined as follows: 
\begin{equation}
\label{stocmatrix}
[\Gamma]_{ji} = \beta_{ij}-\alpha_{ij},
\end{equation}
for all $i \in \mathcal{R}$ and all $j \in \mathcal{S}$ 
(notice the reversal of indices).
This will be later used in order to write down the
differential equation associated to the chemical reaction network.
Notice that we allow $\Gamma$ to have columns which differ only by their sign;
this happens when there are reversible reactions in the network. 

We discuss now how the speed of reactions is affected by the concentrations
of the different species. 
Each chemical reaction takes place continuously in time with its own rate
which is assumed to be only a function of the concentration of the species
taking part in it. 
In order to make this more precise, we 
define the vector $S = [S_1, S_2, \ldots S_{n_s} ]^{\prime}$ of species 
concentrations and, as a function of it, the vector of reaction rates
\[
R(S):= [R_1 (S), R_2 (S), \ldots R_{n_r} (S)]^\prime \,. 
\]
Each reaction rate $R_i$ is a real-analytic function defined on an open set
which contains the non-negative orthant
$\mathcal{O}_+ = \R^{n_s}_{\geq0}$ of $\R^{n_s}$,
and we assume that each $R_i$ depends only on its respective reactants.
(Imposing real-analyticity, that is to say, that the function $R_i$ can be
locally expanded into a convergent power series around each point in its
domain, is a very mild assumption, verified in basically all applications in
chemistry, and it allows stronger statements to be made.)
Furthermore, we assume that each $R_i$ satisfies the following monotonicity
conditions:
 \begin{equation}
 \label{monotonerate1}
 \frac{\partial R_i ( S ) }{ \partial S_j } = \left \{
 \begin{array}{cl}
 \geq 0 & \textrm{if } \alpha_{ij} > 0 \\
 = 0    & \textrm{if } \alpha_{ij} = 0 .
 \end{array}
 \right .
 \end{equation}
We also assume that, whenever the concentration of any of the reactants of a
given reaction is $0$, then, the corresponding reaction does not take place,
meaning that the reaction rate is $0$.  
In other words, if $S_{i_1}, \ldots , S_{i_N}$ are the reactants of reaction
$j$, then we ask that
\[
R_j (S) = 0 \mbox{ for all }
S \mbox{ such that }
[ S_{i_1}, \ldots, S_{i_N} ] \in \partial \mathcal{O}_+ \,,
\]
where $\partial \mathcal{O}_+=\partial \R^N_{\geq 0}$ is the boundary of
$\mathcal{O}_+$ in $\R^N$.
Conversely, we assume that reactions take place if reactants are
available, that is:
\[
R_j (S)>0 \mbox{ whenever } S \mbox{ is such that }
[S_{i_1}, \ldots, S_{i_N} ] \in \textrm{int} [ \R^N_{\geq 0} ]\,.
\]

A special case of reactions is as follows.  One says that a chemical reaction
network is equipped with \emph{mass-action kinetics} if
\[
R_i(S) = k_i \prod_{j=1}^{n_s} S_j^{\alpha_{ij} }
\mbox{ for all } i=1,\ldots,n_r \,.
\]
This is a commonly used form for the functions $R_i(s)$ and amounts to asking
that the reaction rate of each reaction is proportional to the concentration
of each of its participating reactants.
The results in this paper do not require this assumption; in a paper in
preparation will we specialize and tighten our
results when applied to systems with mass-action kinetics.

With the above notations, the chemical reaction network is described by the
following system of differential equations: 
\begin{equation}
\label{chemreactionnetwork}
\dot{S} = \Gamma \, R(S).
\end{equation}
with $S$ evolving in $\mathcal{O}_+$ and where
$\Gamma$ is the stoichiometry matrix.

There are several additional notions useful when analyzing CRN's.
One of them is the notion of a \emph{complex}.
We associate to the network (\ref{list}) a set of complexes,
$C_i$'s, with $i \in \{ 1,2, \ldots, n_c \}$.  Each complex is an integer
combination of species, specifically of the species appearing either as
products or reactants of the reactions in (\ref{list}).
We introduce the following matrix $\tilde{\Gamma}$ as follows:
\[
\tilde{\Gamma} = \left [ \begin{array}{cccccccc} \alpha_{11} &  \alpha_{21}
& \ldots & \alpha_{n_r 1} & \beta_{11} & \beta_{21} &
\ldots & \beta_{n_r 1} \\ 
\alpha_{12} & \alpha_{22} & \ldots & \alpha_{n_r 2} & \beta_{12} & \beta_{22} &
\ldots & \beta_{n_r 2} \\
\vdots & \vdots & & \vdots & \vdots & \vdots & & \vdots \\
\alpha_{1 n_s} & \alpha_{2 n_s} & \ldots & \alpha_{n_r n_s} &
\beta_{1 n_s} & \beta_{2 n_s} &\ldots & \beta_{n_r n_s}
\end{array} 
\right ] 
\]
Then, a matrix representing the complexes as columns can be obtained by
deleting from  $\tilde{\Gamma}$ repeated columns, leaving just one instance of
each; we denote by $\Gamma_c \in\R^{n_s \times n_c}$ the matrix which is thus
constructed.  Each of the columns of $\Gamma_c$ is then associated with a
complex of the network.
We may now associate to each chemical reaction network, a directed graph
(which we call the \emph{C-graph}), whose nodes are the complexes and whose
edges are associated to the reactions (\ref{list}).  An edge $(C_i,C_j)$ is
in the C-graph if and only if $C_i \rightarrow C_j$ is
a reaction of the network.  Note that the C-graph need not be connected
(the C-graph is connected if for any pair of distinct nodes in the graph there
is an undirected path linking the nodes), and lack of connectivity 
cannot be avoided in the analysis.  (This is in contrast with many other
graphs in chemical reaction theory, which can be assumed to be connected
without loss of generality.)  
In general, the C-graph will have several connected components
(equivalence classes under 
the equivalence relation ``is linked by an
undirected path to'', defined on the set of nodes of the graph).

Let $\mathcal{I}$ be the incidence matrix of the C-graph, namely the matrix
whose columns are in one-to-one correspondence with the edges (reactions) of
the graph and whose rows are in one-to-one correspondence with the nodes
(complexes).  
Each column contains a $-1$ in the $i$-th entry and a $+1$ in the $j$-th entry
(and  zeroes in all remaining entries) whenever $(C_i,C_j)$
is an edge of the C-graph (equivalently, when 
$C_i \rightarrow C_j$ is a reaction of the network).
With this notations, we have the following formula, to be used later:
\be{gamma and gamma_c}
\Gamma \;=\; \Gamma_c \,\mathcal{I}\,.
\ee

We denote solutions of (\ref{chemreactionnetwork}) as follows:
$S(t) = \varphi (t,S_0)$, where $S_0 \in \mathcal{O}_+$ is the initial
concentration of chemical species.
As usual in the study of the qualitative behavior of dynamical systems, we
will make use of $\omega$-limit 
sets, which capture the long-term behavior of a
system and are defined as follows:
 \begin{equation}
 \label{omegalimitset}
 \omega(S_0) := \{ S \in \mathcal{O}_+: \varphi (t_n,S_0) \rightarrow S
\textrm{ for some } t_n \rightarrow + \infty \}
 \end{equation}
(implicitly, when talking about $\omega(S_0)$, we assume that $\varphi(t,S_0)$
is defined for all $t\geq0$ for the initial condition $S_0$).
We will be interested in asking whether or not a chemical reaction network
admits solutions in which one or more of the chemical compounds become
arbitrarily small.  The following definition, borrowed from the ecology
literature, captures this intuitive idea.

\bd{persistent}
A chemical reaction network (\ref{chemreactionnetwork}) is
\emph{persistent} if
$\omega (S_0) \cap \partial \mathcal{O}_+ = \emptyset$
for each $S_0 \in \textrm{int} ( \mathcal{O}_+)$.
\ed

We will derive conditions for persistence of general chemical
reaction networks.
Our conditions will be formulated in the language
of Petri nets; these are discrete-event systems equipped with an algebraic
structure that reflects the list of chemical reactions present in the network
being studied, and are defined as follows.
 
\section{Petri Nets}
\label{Petri Nets}

We associate to a CRN a bipartite directed graph (i.e., a directed graph with
two types of nodes) with weighted edges, called the
\emph{species-reaction Petri net}, or SR-net for short.
Mathematically, this is a quadruple
\[
( V_S, V_R, E, W)\,,
\]
where $V_S$ is a finite set of nodes each one associated to a species, $V_R$
is a finite set of nodes (disjoint from $V_S$), each one corresponding to a
reaction, and $E$ is a set of edges as described below.
(We often write $S$ or $V_S$ interchangeably, or $R$ instead of $V_R$, by
identifying species or reactions with their respective indices; the context
should make the meaning clear.)
The set of all nodes is also denoted by $V \doteq V_R \cup V_S$.

The edge set $E\subset V \times V$ is defined as follows.
Whenever a certain reaction $R_i$ belongs to the CRN:
\begin{equation}
\label{arbitrarydirection}
 \sum_{j \in \mathcal{S} } \alpha_{ij} S_j \quad \rightarrow 
 \quad \sum_{j \in \mathcal{S} } \beta_{ij} S_j \,,
\end{equation}
we draw an edge from $S_j \in V_S$ to $R_i \in V_R$ for all 
$S_j$'s such that $\alpha_{ij} >0$.  That is, $(S_j,R_i) \in E$ iff
$\alpha_{ij}>0$,
and we say in this case that $R_i$ is an \emph{output reaction for} $S_j$.
Similarly, we draw an edge from $R_i \in V_R$ to every
$S_j \in V_S$ such that $\beta_{ij}>0$.
That is, $(R_i,S_j) \in E$ whenever $\beta_{ij}>0$,
and we say in this case that $R_i$ is an \emph{input reaction for} $S_j$.

Notice that edges only connect species to reactions and vice versa, but never
connect two species or two reactions.

The last element to fully define the Petri net is the function
$W:E\rightarrow\N$, which associates to each edge   
a positive integer according to the rule:
\[
W( S_j,R_i ) = \alpha_{ij} \quad\mbox{ and }\quad W(R_i, S_j) = \beta_{ij} \,.
\]

Several other definitions which are commonly used in the Petri net literature
will be of interest in the following. 
We say that a row or column vector $v$ is non-negative, and we denote it by
$v \succeq 0$ if it is so entry-wise.
We write $v \succ 0$ if $v \succeq 0$ and $v \neq 0$. 
A stronger notion is instead $v \gg 0$, which indicates $v_i > 0$ for all $i$.

\bd{P-semiflow}
A \emph{P-semiflow} is any row vector $c \succ 0$ such that $c \, \Gamma = 0$.
Its \emph{support} is the set of indices 
$\{ i \in V_S : c_i > 0\}$.
A Petri net is said to be \emph{conservative} if there exists a
P-semiflow $c \gg 0$.
\ed

Notice that P-semiflows for the system (\ref{chemreactionnetwork}) correspond
to non-negative linear first integrals, that is, linear functions $S\mapsto cS$
such that $(d/dt)cS(t)\equiv0$ along all solutions
of~(\ref{chemreactionnetwork})
(assuming that the span of the image of $R(S)$ is $\R^{n_r}$).
In particular, a Petri net is conservative if and only if there is a positive
linear conserved quantity for the system.
(Petri net theory views Petri nets as ``token-passing'' systems, and, in
that context, P-semiflows, also 
called \emph{place-invariants}, amount
to conservation relations for the ``place markings'' of the network, that show
how many tokens there are in each ``place,'' the nodes associated to species
in SR-nets.  We do not make use of this interpretation in this paper.)

\bd{T-semiflow}
A \emph{T-semiflow} is any column vector $v \succ 0$ such that $\Gamma \, v=0$.
A Petri net is said to be \emph{consistent} if there exists a T-semiflow
$v \gg 0$.
\ed

The notion of T-semiflow corresponds to the existence of a collection of
positive reaction rates which do not produce any variation in the
concentrations of the species.  In other words, $v$ can be viewed as a
set of \emph{fluxes} that is in equilibrium (\cite{schuster}).
(In Petri net theory, the terminology is ``T-invariant,'' and the fluxes are 
flows of tokens.)

A chemical reaction network is said to be \emph{reversible} if each chemical
reaction has an inverse reaction which is also part of the network.
Biochemical models are most often non-reversible.  For this reason, a far
milder notion was
introduced~\cite{hornjackson1,hornjackson2,feinberg0,feinberg1,feinberg2}:
A chemical reaction network is said to be \emph{weakly reversible} if
each connected component of the C-graph is strongly connected (meaning that
there is a directed path between any pair of nodes in each connected 
component).
In algebraic terms, weak reversibility amounts to existence of $v \gg 0$ such
that $\mathcal{I} v = 0$ (see Corollary 4.2 of \cite{feinberg}),
so that in particular, using~(\ref{gamma and gamma_c}),
also $\Gamma v = \Gamma_c  \mathcal{I} v = 0$.
Hence a chemical reaction network that is weakly reversible
has a consistent associated Petri net.

A few more definitions are needed in order to state our main results.

\bd{siphon}
A 
nonempty set $\Sigma \subset V_S$ is called a \emph{siphon} 
if each input reaction associated to $\Sigma$ is also an output reaction
associated to $\Sigma$.
A siphon is a \emph{deadlock} if its set of output reactions is all of $V_R$.
A deadlock is \emph{minimal} if it does not contain (strictly) any other
deadlocks.
A pair of distinct deadlocks $\Sigma_1$ and $\Sigma_2$ is said to be
\emph{nested} if either $\Sigma_1\subset \Sigma_2$ or $\Sigma_2\subset
\Sigma_1$.
\ed
For later use we associate a particular set to a siphon $\Sigma$ as follows:
$$
L_\Sigma=\{x\in \mathcal{O}_+\,|\, x_i=0 \Longleftrightarrow i\in \Sigma\}.
$$
It is also useful to introduce a binary relation ``reacts to'', which we denote by
$\rightarrowtail$, and we define as follows: $S_i \rightarrowtail S_j$
whenever there exists a chemical reaction $\mathcal{R}_k$, so that 
\[
\sum_{l  \in \mathcal{S}} \alpha_{kl} S_l \rightarrow \sum_{l \in \mathcal{S}}
   \beta_{kl} S_l
\]
with $\alpha_{ki}>0$, $\beta_{kj}>0$.
If the reaction number is important, we also write
\[
S_i \rightarrowtail^{ \hspace{-3mm} k} S_j
\]
(where $k \in \mathcal{R}$).
With this notation, the notion of siphon can be rephrased as follows:
$Z \subset \mathcal{S}$ is a siphon for a chemical reaction network if
for every $S \in Z$ and $k \in \mathcal{R}$ such that 
$\tilde{S}_k := \{ T \in \mathcal{S} : T \rightarrowtail^{ \hspace{-3mm} k} S \} \neq \emptyset$,  
it holds $\tilde{S}_k  \cap Z \neq \emptyset$. 

\section{Necessary conditions}
\label{Necessary conditions}

Our first result will relate persistence of a chemical reaction network to
consistency of the associated Petri net. 

\bt{Theorem 1}
Let (\ref{chemreactionnetwork}) be the equation describing the time-evolution
of a conservative and persistent chemical reaction network. Then, the
associated Petri net is consistent. 
\ets

\bpr
Let $S_0 \in \textrm{int} ( \mathcal{O}_+ )$ be any initial condition.
By conservativity, solutions satisfy $cS(t)\equiv cS_0$, and hence remain
bounded, and therefore $\omega(S_0)$ is a nonempty compact set.
Moreover, by persistence, 
$\omega(S_0) \cap \partial \mathcal{O}_+ = \emptyset$,
so that $R( \tilde{S}_0 ) \gg 0$,
for all $\tilde{ S}_0 \in \omega(S_0)$.
In particular, by compactness of $\omega (S_0)$ and continuity of $R$, there
exists a positive vector $v \gg 0$, so that
\[
R( \tilde{S}_0 ) \succeq v \;\;\mbox{for all}\;\; \tilde{S}_0 \in \omega (S_0)
\;.
\]
Take any $\tilde{S}_0 \in \omega (S_0)$. By invariance of $\omega(S_0)$, we
have $R ( \varphi (t, \tilde{S}_0 ) ) \succeq v$ for all $t \in \R$.
Consequently, taking asymptotic time averages, we obtain:
\begin{equation}
\label{average}
 0 = \lim_{T \rightarrow + \infty } 
 \frac{ \varphi(T, \tilde{S}_0 ) - \tilde{S}_0}{T} = 
 \lim_{T \rightarrow + \infty} \frac{1}{T} 
 \int_0^T \Gamma R( \varphi(t, \tilde{S}_0 ) ) \, dt
\end{equation}
(the left-hand limit is zero because $\varphi(T, \tilde{S}_0 )$ is bounded).
However,  
\[
\frac{1}{T} \int_0^T  R( \varphi(t, \tilde{S}_0 ) ) \, dt \succeq v  
\] 
for all $T>0$. Therefore, taking any subsequence $T_n \rightarrow + \infty$ so
that there is a finite limit:
\[ 
\lim_{n \rightarrow + \infty} \frac{1}{T_n} \int_0^{T_n}  R( \varphi(t,
\tilde{S}_0 ) ) \, dt = \bar{v} \succeq v \,.
\]
We obtain, by virtue of (\ref{average}), that $\Gamma \, \bar{v} = 0$. This
completes the proof of consistency, since $\bar{v} \gg 0$.
\epr

\section{Sufficient Conditions}
\label{Sufficient Conditions}

In this present Section, we derive sufficient conditions for insuring
persistence of a chemical reaction network on the basis of Petri net
properties.

\bt{Theorem 2}
Consider a chemical reaction network satisfying the following assumptions:
\begin{enumerate}
\item
its associated Petri net is conservative;
\item
each siphon contains the support of a P-semiflow.
\end{enumerate}
Then, the network is persistent.
\ets

We first prove a number of technical results.
The following general fact about differential equations will be useful.

For each real number $p$, let $\sign p:=1,0,-1$ if $p>0$, $p=0$,
or $p<0$ respectively, and for each vector $x=(x_1,\ldots ,x_n)$, 
let $\sign x:=(\sign x_1,\ldots ,\sign x_n)'$.
When $x$ belongs to the closed positive orthant $\Rp^n$,
$\sign x\in \{0,1\}^n$.

\bl{Lemma eds}
Let $f$ be a real-analytic vector field defined on some open neighborhood
of $\Rpn$, and suppose that $\Rpn$ is forward invariant for the flow of $f$.
Consider any solution 
$\barx(t)$ of $\dot x=f(x)$, evolving in $\Rpn$ and defined
on some open interval $J$.
Then, $\sign \barx(t)$ is constant on $J$.
\els

\bpr
Pick such a solution, and define
\[
Z := \, \{i \,|\, \barx_i(t) = 0 \,\mbox{ for all }\, t\in J \}\,.
\]
Relabeling variables if necessary, we assume without loss of generality
that $Z=\{r+1,\ldots ,n\}$, with $0\leq r\leq n$, and we write equations in the
following block form:
\beqn
\dot y &=& g(y,z)\\
\dot z &=& h(y,z)
\eeqn
where $x'=(y',z')'$ and $y(t)\in \R^r$, $z(t)\in \R^{n-r}$.
(The extreme cases $r=0$ and $r=n$ correspond to $x=z$ and $x=y$
respectively.)
In particular, we write 
$\barx'=(\bary',\barz')'$ for the trajectory of interest.
By construction, $\barz\equiv 0$, and 
the sets
\[
B_i:=\{t\,|\,\bary_i(t)=0\}
\]
are proper subsets of $J$,
for each $i\in \{1,\ldots ,r\}$.
Since the vector field is real-analytic, each coordinate function
$\bary_i$ is real-analytic (see e.g. \cite{mct}, Proposition C.3.12),
so, by the principle of analytic continuation, each $B_i$ is a discrete set.  
It follows that 
\[
G:=J\setminus\bigcup _{i=1}^r B_i
\]
is an (open) dense set, and
for each $t\in G$, $\bary(t)\in \mbox{inter}\,\Rp^r$,
the interior of the positive orthant.

We now consider the following system on $\R^r$:
\[
\dot y = g(y,0)\,.
\]
This is again a real-analytic system, and $\Rp^r$ is forward invariant.
To prove this last assertion, note that forward invariance of the closed
positive orthant is equivalent to the following property:
\begin{center}
for any $y\in \Rp^r$ and any $i\in \{1,\ldots ,r\}$ such that $y_i=0$, 
$g_i(y,0)\geq 0$.
\end{center}
Since $\Rp^n$ is forward invariant for the original system, we know, by
the same property applied to that system, that 
for any $(y,z)\in \Rp^n$ and any $i\in \{1,\ldots ,r\}$ such that $y_i=0$, 
$g_i(y,z)\geq 0$.  Thus, the required property holds (case $z=0$).
In particular, $\mbox{inter}\,\Rp^r$
is also forward invariant (see e.g.~\cite{our1}, Lemma III.6).
By construction, $\bary$ is a solution of $\dot y = g(y,0)$,
$\bary(t)\in \mbox{inter}\,\Rp^r$ for each $t\in G$, 
Since $G$ is dense and $\mbox{inter}\,\Rp^r$ is forward invariant,
it follows that $\bary(t)\in \mbox{inter}\,\Rp^r$ for all $t\in J$.
Therefore,
\[
\sign\barx(t)=
(1_r,0_{n-r})'\;\;\mbox{for all}\;\;
t\in J\,
\]
where $1_r$ is a vector of $r$ $1$'s and
$0_{n-r}$ is a vector of $n-r$ $0$'s.
\epr


We then have an immediate corollary:

\bl{Lemma 2}
Suppose that $\Omega \subset \mathcal{O}_+$ is a closed set, invariant for
(\ref{chemreactionnetwork}).
Suppose that $\Omega \cap L_Z$ is non-empty,
for some $Z \subset \mathcal{S}$.
Then, $\Omega \cap  L_Z $ is also invariant with respect to
(\ref{chemreactionnetwork}). 
\els

\bpr
Pick any $S_0\in\Omega \cap L_Z$.  By invariance of $\Omega$, the solution
$\varphi(t,S_0)$ belongs to $\Omega$ for all $t$ in its open domain of
definition $J$, so, in particular (this is the key fact),
$\varphi(t,S_0)\in\mathcal{O}_+$ for all $t$ (negative as well as positive).
Therefore, it also belongs to $L_Z$, since its
sign is constant by \reflemma{Lemma eds}.
\epr

In what follows, we will make use of the Bouligand tangent cone
$TC_{\xi} (K)$ of a set $K\subset \mathcal{O}_+$ at
a point $\xi\in \mathcal{O}_+$, defined as follows:
\[
TC_{\xi} (K) \;=\; \left\{ v\in\R^n: \exists   k_n  \in K, 
  k_n \rightarrow \xi \, \textrm{and} \, \lambda_n \searrow 0:
  \frac{1}{\lambda_n} (k_n - \xi) \rightarrow v  \right\} \,. 
\]
Bouligand cones provide a simple criterion to check forward invariance of
closed sets (see e.g.~\cite{cellina}): a closed set $K$ is
forward invariant for (\ref{chemreactionnetwork}) if and only if 
$\Gamma R(\xi)\in TC_{\xi} (K)$ for all $\xi \in K$.
However, below we consider a condition involving tangent cones to the sets
$L_Z$, which are not closed.
Note that, for all index sets $Z$ and all points $\xi$ in $L_Z$,
\[
TC_{\xi} \left(L_Z\right) \;=\;
          \left\{v\in\R^n : v_i=0\;\forall\,i\in Z\right\}\,.
\]

\bl{Lemma 3}
Let  $Z \subset \mathcal{S}$ be non-empty and  $\xi \in L_Z$ be such that
$\Gamma R( \xi ) \in TC_{\xi} ( L_Z )$. Then $Z$ is a \emph{siphon}.
\els

\bpr
By assumption $\Gamma R( \xi ) \in TC_{\xi} (L_Z)$ for some $\xi \in L_Z$.
This implies that $[ \Gamma R ( \xi ) ]_i = 0 $ for all $i \in Z$. Since
$\xi_i = 0$ for all $i \in Z$, all reactions in which $S_i$ is involved as a
reactant are shut off at $\xi$; hence, the only possibility for 
$[\Gamma R(\xi ) ]_i = 0$ is that all reactions in which $S_i$ is
involved as a product are also shut-off.
Hence, for all $k \in \mathcal{R}$, and all $l
\in \mathcal{S}$ so that $S_l \rightarrowtail^{ \hspace{-3mm} k} S_i$, we
necessarily have that $R_k ( \xi )=0$.

Hence, for all $k \in \mathcal{R}$ so that
$\tilde{S}_k = \{ l \in \mathcal{S}: 
    S_l \rightarrowtail^{  \hspace{-3mm} k} S_i \}$
is non-empty, there 
must exist an  $l \in \tilde{S}_k$ so that $\xi_l = 0$. 
But then necessarily, $l \in Z$, 
showing that $Z$ is indeed a siphon.
\epr

The above Lemmas are instrumental to proving the following Proposition:

\bp{Proposition 1}
Let $\xi \in \mathcal{O}^+$ be such that
$\omega(\xi) \cap L_Z \neq \emptyset$ for some $Z \subset \mathcal{S}$. Then
$Z$ is a siphon.
\eps

\bpr
Let $\Omega $ be the closed and invariant set $\omega(\xi)$.
Thus, by \reflemma{Lemma 2}, 
the non-empty set $L_Z \cap \Omega$ is also invariant. Notice that 
\[
\textrm{cl} [ L_Z ] \;=\; \bigcup_{W \supseteq Z} L_W \,.
\]
Moreover, $L_W \cap \Omega$ is invariant for all $W \subset \mathcal{S}$ such
that $L_W \cap \Omega$ is non-empty.
Hence,
\[
\textrm{cl}[ L_Z ]\cap \Omega \;= \;\bigcup_{W \supseteq Z} [ L_W \cap\Omega]
\]
is also invariant.
By the characterization of invariance for closed sets in terms of Bouligand
tangent cones, 
we know that, for any $\eta \in \textrm{cl} [L_Z] \cap \Omega$ we have
\[
\Gamma R(\eta) \;\in\; TC_{\eta} ( \Omega \cap \textrm{cl}(L_Z))\;\subset\;
  TC_{\eta} ( \textrm{cl} ( L_Z ) ) \,.
\]
In particular, for $\eta \in L_Z \cap \Omega$ (which by assumption exists), 
$\Gamma R ( \eta) \in TC_{\eta} ( L_Z )$ so that, by
virtue of \reflemma{Lemma 3} we may conclude $Z$ is a siphon.
\epr

Although at this point \refprop{Proposition 1} would be enough
to prove \reftheorem{Theorem 2}, it
is useful to clarify the meaning of the concept of a ``siphon'' here.
It hints at the fact, made precise in the Proposition below, that removing all
the species of a siphon from the network (or equivalently setting their
initial concentrations equal to $0$) will prevent those species from being
present at all future times. Hence, those species literally ``lock'' a part of
the network and shut off all the reactions that are therein involved. In
particular, once emptied a siphon will never be full again.  This explains why
a siphon is sometimes also called a ``locking set'' in the Petri net
literature.  A precise statement of the foregoing remarks is as follows.

\bp{Proposition 2}
Let $Z\subset \mathcal{S}$ be non-empty. Then $Z$ is a siphon if and only if
$\textrm{cl}(L_Z)$ is forward invariant for $(\ref{chemreactionnetwork})$.
\eps

\bpr
\emph{Sufficiency:}
Pick $\xi \in L_Z\neq \emptyset$. Then forward invariance of
$\textrm{cl}(L_Z)$ implies that 
$\Gamma R(\xi)\in TC_{\xi}(\textrm{cl}(L_Z))=TC_{\xi}(L_Z)$, where the last
equality holds since 
$\xi\in L_Z$. 
It follows from \reflemma{Lemma 3} that $Z$ is a siphon.

\emph{Necessity:}
Pick $\xi \in \textrm{cl}(L_Z)$. This implies that 
$\xi_i=0$ for all $i\in Z\cup Z'$, where $Z'\subset \mathcal{S}$ could be
empty. By the characterization 
of forward invariance of closed sets in terms of tangent Bouligand cones, 
it suffices to show that $[\Gamma R(\xi)]_i=0$ for all $i\in Z$, and that 
$[\Gamma R(\xi)]_i\geq 0$ for all $i\in Z'$ whenever 
$Z'\neq \emptyset$. Now by $(\ref{chemreactionnetwork})$,
\begin{equation}
\label{inequality}
[\Gamma R(\xi)]_i\;=\;
\sum_k \beta_{ki} R_k(\xi)-\sum_l \alpha_{li}R_l(\xi)\;=\;
\sum_k \beta_{ki} R_k(\xi)-0\;\geq\; 0\,,
\end{equation}
which already proves the result for $i\in Z'$. Notice that the second sum is
zero because if $\alpha_{li}>0$, then species $i$ is a reactant of reaction
$l$, which implies that $R_l(\xi)=0$ since $\xi_i=0$.
So we assume henceforth that $i\in Z$. We claim that the sum on the right side
of $(\ref{inequality})$ is zero.
This is obvious if the sum is void. If it is non-void, then each term which is such that $\beta_{ki}>0$ must be zero. 
Indeed, for each such term we have that $R_k(\xi)=0$ because $Z$ is a
siphon. This concludes the proof of \refprop{Proposition 1}.
\epr

\subsection*{Proof of Theorem~\protect\ref{Theorem 2}}

Let $\xi \in \textrm{int} ( \mathcal{O}_+)$ be arbitrary and let $\Omega$
denote the corresponding $\omega$-limit set $\Omega = \omega (\xi )$. We claim
that the intersection of $\Omega$ and the boundary of $\mathcal{O}_+$ is
empty.

Indeed, suppose that the intersection is nonemty.  Then, $\Omega$ would
intersect $L_Z$, for some $\emptyset \neq Z \subset \mathcal{S}$. In
particular, by \refprop{Proposition 1}, $Z$ would be a siphon.  Then, by our
second assumption, there
exists a 
non-negative first integral $c S$, whose support is included in
$Z$, so that necessarily $c S (t_n,\xi) \rightarrow 0$ at least along
a suitable sequence $t_n \rightarrow + \infty$.  
However, $c S(t,\xi) = c \xi > 0$
for all $t \geq 0$, thus giving a contradiction.
\qed

\section{Applications}
\label{Applications}

We now apply our results to obtain persistence results for variants of the
reaction (b) shown in Figure~\ref{futile_cycle} as well as for cascades of
such reactions.

\subsection{Example 1}

\begin{figure}
\centerline{\includegraphics[width=10cm]{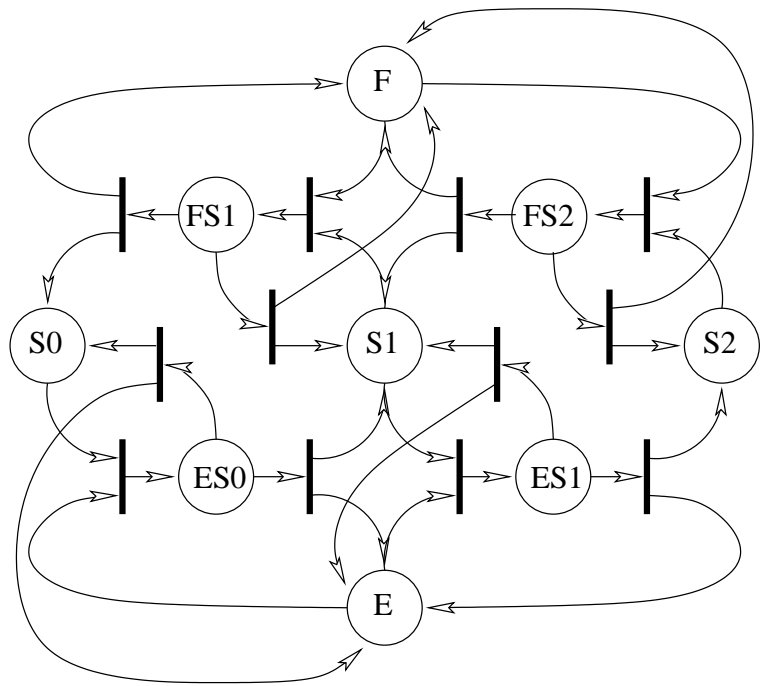}}
\caption{Petri net associated to reactions (\ref{mapk}).}
\label{mapkstage}
\end{figure}

We first study reaction~(\ref{mapk}).
Note that reversible reactions were denoted by a ``$\leftrightarrow$'' in
order to avoid having to rewrite them twice. 
The Petri net associated to (\ref{mapk}) is shown if Fig.~\ref{mapkstage}.
The network comprises nine distinct species,
labeled $S_0$, $S_1$, $S_2$, $E$, $F$, $ES_0$, $ES_1$, $FS_2$, $FS_1$.
It can be verified that the Petri net in Fig.~\ref{mapkstage}
is indeed consistent
(so it satisfies the necessary condition).
To see this, order the species and reactions by the obvious order obtained
when reading $(\ref{mapk})$ from left to right and from top to
bottom (e.g., $S_1$ is the fourth species and the reaction 
$E+S_1\rightarrow ES_1$ is the fourth reaction).
The construction of the matrix $\Gamma$ is now clear, and it 
can be verified that $\Gamma v=0$ with 
$v=[2\, 1\, 1\, 2\, 1\, 1\, 2\, 1\, 1\, 2\, 1\, 1\,]'$. 
The network itself, however, is not weakly reversible,  since neither of the
two connected components of (\ref{mapk}) is strongly connected.
Computations show that there are three minimal siphons:

$\{ E, ES_0, ES_1 \}$,

$\{ F, FS_1, FS_2 \}$, 

\noindent
and

$\{ S_0, S_1,  S_2, ES_0, ES_1, FS_2, FS_1 \}$.

\noindent
Each one of them contains the support of a P-semiflow; in fact there 
are three independent conservation laws:

$E + ES_0 + ES_1 = {\rm const}_1$,

$ F + FS_2 + FS_1 = {\rm const}_2$, 
\noindent
and

$S_0 + S_1 + S_2 + ES_0 + ES_1 + FS_2 + FS_1 = {\rm const}_3$,

\noindent
whose supports coincide with the three mentioned siphons. 
Since the sum of these three conservation laws is also a conservation law, the
network is conservative.
Therefore, application of \reftheorem{Theorem 2} guarantees that the network is indeed
persistent.

\subsection{Example 2}

\begin{figure}
\centerline{\includegraphics[width=14cm]{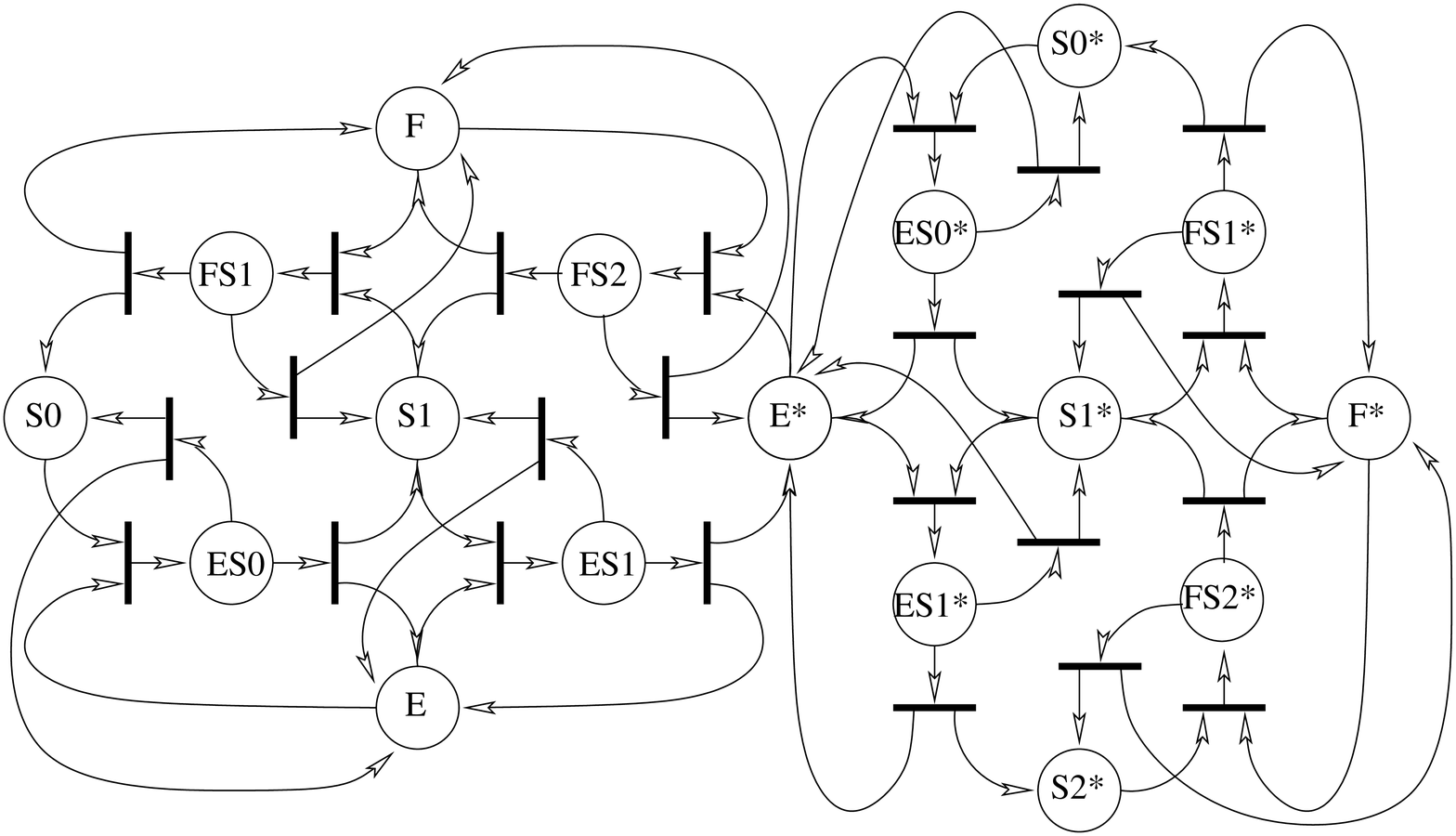}}
\caption{Petri net associated to reactions (\ref{mapk2}).}
\label{mapkstage2}
\end{figure}

As remarked earlier, examples as the above one are often parts of cascades in
which the product (in MAPK cascades, a doubly-phosphorilated species) $S_2$ in
turn acts as an enzyme for the 
following stage. 
One model with two stages is as follows (writing $S_2$ as
$E^\star$ in order to emphasize its role as a kinase for the subsequent stage):
\begin{equation}
\label{mapk2}
\begin{array}{ccccccccc}
E + S_0 &\leftrightarrow& ES_0 & \rightarrow & E + S_1 &
\leftrightarrow & ES_1 & \rightarrow &E + E^\star \\
F + E^{\star} & \leftrightarrow& F S_2 & \rightarrow & F + S_1 & \leftrightarrow &
F S_1 & \rightarrow &F + S_0 \\
E^{\star} + S_0^{\star} &\leftrightarrow& ES_0^{\star} & \rightarrow & E^{\star} + S_1^{\star} &
\leftrightarrow & ES_1^{\star} & \rightarrow &E^{\star} + S_2^{\star} \\
F^{\star} + S_2^{\star} & \leftrightarrow& F S_2^{\star} & \rightarrow & F^{\star} + S_1^{\star} & \leftrightarrow &
F S_1^{\star} & \rightarrow &F^{\star} + S_0^{\star}. \\
\end{array}
\end{equation}
The overall reaction is shown in Fig. \ref{mapkstage2}.
Note -- using the labeling of species and reaction as in the previous
example -- that 
$\Gamma v=0$ with $v=[v_1' \, v_1' \, v_1' \, v_1']'$ and $v_1=[2\, 1\, 1\, 2\, 1\, 1]'$, and hence the network is 
consistent. 
There are five minimal siphons for this network, namely:

$\{ E, ES_0, ES_1 \}$,

$\{ F, FS_2, FS_1 \}$,

$\{ F^{\star}, FS_2^{\star}, FS_1^{\star} \}$,

$\{ S_0^{\star}, S_1^{\star}, S_2^{\star}, ES_0^{\star}, ES_1^{\star}, FS_2^{\star}, FS_1^{\star} \}$,

\noindent
and

$\{ S_0, S_1, E^{\star}, ES_0, ES_1, FS_2, FS_1, ES_0^{\star},ES_1^{\star} \}$.

\noindent
Each one of them is the 
support of a P-semiflow, and there are five conservation laws:

$ E+ES_0+ES_1={\rm const}_1$,

$F+FS_2+FS_1={\rm const}_2$,

$F^{\star}+FS_2^{\star}+FS_1^{\star}={\rm const}_3$,

$S_0^{\star}+S_1^{\star}+S_2^{\star}+ES_0^{\star}+ES_1^{\star}+FS_2^{\star}+FS_1^{\star}={\rm const}_4$, 

\noindent
and

$S_0+S_1+E^{\star}+ES_0+ES_1+FS_2+FS_1+ES_0^{\star}+ES_1^{\star}={\rm const}_5$.

\noindent
As in the previous example, the network is conservative since the sum of these
conservation laws is also a conservation law. 
Therefore the overall network is persistent, by virtue 
of~\reftheorem{Theorem 2}. 

\subsection{Example 3}

\begin{figure}
\centerline{\includegraphics[width=14cm]{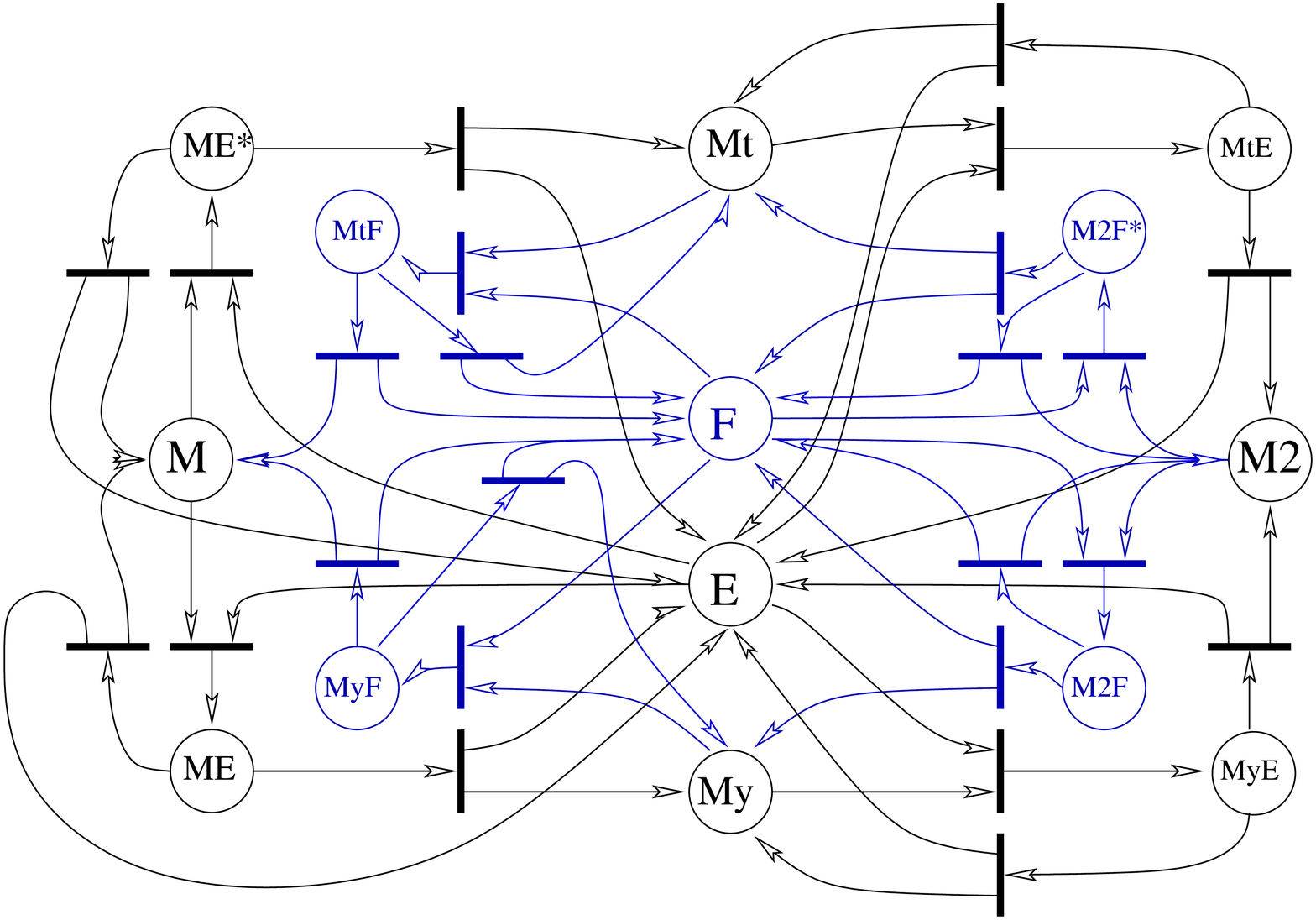}}
\caption{Petri net associated to the network (\ref{doublesite}).}
\label{doublesitefig}
\end{figure}

An alternative mechanism for dual phosphorilation in MAPK cascades, considered
in~\cite{kholodenko}, differs
from the previous ones in that it becomes relevant in what order the two
phosphorylations occur.
(These take place at two different sites, a threonine and
a tyrosine residue).  The corresponding network can be modeled as follows:
\begin{equation}
\label{doublesite}
\begin{array}{ccccccccc} M + E &\leftrightarrow& ME& \rightarrow& M_y + E &\leftrightarrow& M_y E &\rightarrow & M_2 + E \\
M+E  & \leftrightarrow &
ME^{\star} & \rightarrow & M_t + E & \leftrightarrow & M_t E & \rightarrow & M_2 + E \\
M_2 + F &\leftrightarrow& M_2F& \rightarrow& M_y + F &\leftrightarrow& M_y F &\rightarrow & M + F \\
M_2 + F & \leftrightarrow &
M_2F^{\star} & \rightarrow & M_t + F & \leftrightarrow & M_t F & \rightarrow & M + F. \\
\end{array}
\end{equation}
See Fig. \ref{doublesitefig} for the corresponding Petri net.
This network is consistent. 
Indeed, $\Gamma v=0$ for the same $v$ as in the previous example. 
Moreover it admits three siphons of minimal support:

$\{ E, ME, ME^{\star}, M_y E, M_t E \}$,

$\{ F, M_y F, M_t F, M_2 F, M_2 F^{\star} \}$, 

\noindent
and

$\{ M, ME, ME^{\star}, M_y, M_t, M_y E, M_t E, M_2,M_2 F, M_2 F^{\star}, M_t F, M_y F \}$.

\noindent
Each of them is also the support of a conservation law, respectively for $M$,$E$ and $F$ molecules. 
The sum of these conservation laws, is also a conservation law and therefore the network is conservative. 
Thus the \reftheorem{Theorem 2} again applies and the network is persistent.


\clearpage
\newpage
\begin{thebibliography}{99}

\newcommand{\book}[1]{{\em #1\/},}
\newcommand{\inbook}[1]{in {\em #1\/},}
\newcommand{\journal}[1]{{\em #1\/}}
\newcommand{\jvol}[1]{{\bf #1}}
\newcommand{\jyear}[1]{(#1),}
\newcommand{\pp}[1]{pp.\ #1.}
\newcommand{\papertitle}[1]{``#1''}
\newcommand{\EDS}{\AUTHOR{E.D.}{Sontag} }
\newcommand{\AUTHOR}[2]{#1 #2,}    
\newcommand{\STARTWITH}{}
\newcommand{\ENDWITH}{}


\bibitem{cdc06}
\AUTHOR{D}{Angeli}
\AUTHOR{P}{De Leenheer}
\AUTHOR{E.D.}{Sontag}
\papertitle{On the structural monotonicity of chemical reaction networks}
{Proc.\ IEEE Conf.\ Decision and Control, San Diego, Dec.\ 2006}, IEEE Publications,
\jyear{2006}
to appear.

\bibitem{our1}
\AUTHOR{D.}{Angeli}
\AUTHOR{E.D.}{Sontag}
\papertitle{Monotone control systems}
\journal{IEEE Trans.\ Autom.\ Control}
\jvol{48}
\jyear{2003}
\pp{1684--1698}

\bibitem{our3}
\AUTHOR{D.}{Angeli}
\AUTHOR{J.E.}{Ferrell, Jr.}
\AUTHOR{E.D.}{Sontag}
\papertitle{Detection of multi-stability, bifurcations, and hysteresis in a large class of biological positive-feedback systems}
\journal{Proceedings of the National Academy of Sciences USA}
\jvol{101}
\jyear{2004}
\pp{1822--1827}

\bibitem{trans_invariance_include_conf}
\AUTHOR{D.}{Angeli}
\AUTHOR{E.D.}{Sontag}
\papertitle{A global convergence result for strongly monotone systems with positive translation invariance}
submitted.
(Summarized version in
``A note on monotone systems with positive translation invariance,''
\emph{Proc. 14th IEEE Mediterranean Conference on Control and Automation,
June 28-30, 2006, Ancona, Italy}\\
{\small \tt http://www.diiga.univpm.it/MED06})

\bibitem{cellina}
J-P. Aubin,
A. Cellina,
\emph{Differential Inclusions: Set-Valued Maps and Viability Theory},
Springer-Verlag, 1984.

\bibitem{bhatia}
N.P. Bhatia,
G.P. Szeg\"o,
\emph{Stability Theory of Dynamical Systems},
Springer-Verlag, Berlin, 1970.

\bibitem{waltman1}
G. Butler,
P. Waltman,
\papertitle{Persistence in dynamical systems}
\journal{J. Differential Equations}
\jvol{63}
\jyear{1986}
\pp{255-263}

\bibitem{waltman2}
G. Butler,
H.I. Freedman,
P. Waltman,
\papertitle{Uniformly persistent systems}
\journal{Proc. Am. Math. Soc.}
\jvol{96}
\jyear{1986}
\pp{425-430}

\bibitem{our4}
\AUTHOR{M.}{Chaves}
\AUTHOR{E.D.}{Sontag}
\AUTHOR{R.J.}{Dinerstein}
\papertitle{Steady-states of receptor-ligand dynamics: A theoretical framework}
\journal{J.\ Theoretical Biology}
\jvol{227}
\jyear{2004}
\pp{413--428}

\bibitem{clarke}
B.L. Clarke,
\papertitle{Stability of complex reaction networks}
\journal{Adv. Chem. Phys.}
\jvol{43}
\jyear{1980}
\pp{1-216}

\bibitem{conradi}
C. Conradi,
J. Saez-Rodriguez,
E.-D. Gilles,
J. Raisch
\papertitle{Using chemical reaction network theory to discard a kinetic mechanism hypothesis}
in
\emph{Proc. FOSBE 2005 (Foundations of Systems Biology in Engineering), Santa Barbara, Aug. 2005}.
\pp{325-328}.

\bibitem{our2}
\AUTHOR{P.}{De Leenheer}
\AUTHOR{D.}{Angeli}
\AUTHOR{E.D.}{Sontag}
\papertitle{Monotone chemical reaction networks}
\journal{J.\ Mathematical Chemistry}
\jyear{2006}
to appear.

\bibitem{feinberg0}
M. Feinberg,
F.J.M. Horn,
\papertitle{Dynamics of open chemical systems and algebraic structure of underlying reaction network}
\journal{Chemical Engineering Science}
\jvol{29}
\jyear{1974}
\pp{775-787}

\bibitem{feinberg1}
M.\ Feinberg,
``Chemical reaction network structure and the stabiliy of complex isothermal
reactors - I. The deficiency zero and deficiency one theorems,''
Review Article 25, {\em Chemical Engr.\ Sci.} {\bf 42}(1987), pp.\ 2229-2268.

\bibitem{feinberg2}
M.\ Feinberg,
``The existence and uniqueness of steady states for a class
of chemical reaction networks,''
{\em Archive for Rational Mechanics and Analysis}
{\bf 132}(1995), pp.\ 311-370.

\bibitem{feinberg}
M. Feinberg,
\papertitle{Lectures on chemical reaction networks}
\journal{Lectures at the Mathematics Research Center, University of Wisconsin},
1979.\\
{\small \tt http://www.che.eng.ohio-state.edu/$\tilde{}$feinberg/LecturesOnReactionNetworks/}

\bibitem{gard}
T.C. Gard,
\papertitle{Persistence in food webs with general interactions}
\journal{Math. Biosci.}
\jvol{51}
\jyear{1980}
\pp{165--174.}

\bibitem{petri1}
H. Genrich,
R. K\"uffner,
K. Voss,
\papertitle{Executable Petri net models for the analysis of metabolic pathways}
\journal{Int.\ J.\ on Software Tools for Technology Transfer (STTT)}
\jvol{3}
\jyear{2001}
\pp{394-404}

\bibitem{hal2}
M. Hirsch,
H.L. Smith,
\title{Monotone dynamical systems}
\inbook{Handbook of Differential Equations, Ordinary Differential Equations (second volume)}
(A. Canada, P. Drabek, and A. Fonda, eds.),
Elsevier,
2005.

\bibitem{hofbauer}
J. Hofbauer,
J.W.-H. So,
\papertitle{Uniform persistence and repellors for maps}
\journal{Proceedings of the American Mathematical Society}
\jvol{107}
\jyear{1989}
\pp{1137-1142}

\bibitem{petri2}
R. Hofest\"adt,
\papertitle{A Petri net application to model metabolic processes}
\journal{Syst. Anal. Mod. Simul.}
\jvol{16}
\jyear{1994}
\pp{113-122}


\bibitem{hornjackson1}
F.J.M.\ Horn,
R.\ Jackson, 
``General mass action kinetics,''
{\em Arch.\ Rational Mech.\ Anal.} {\bf 49}(1972), pp.\ 81-116.

\bibitem{hornjackson2}
F.J.M.\ Horn,
``The dynamics of open reaction systems,''
in {\em Mathematical aspects of chemical and biochemical problems and quantum
chemistry (Proc. SIAM-AMS Sympos.\ Appl.\ Math., New York, 1974)},
pp. 125-137. SIAM-AMS Proceedings, Vol. VIII, Amer.\ Math.\ Soc., Providence,
1974.

\bibitem{ferrell}
C.-Y.F. Huang,
Ferrell, J.E.,
\papertitle{Ultrasensitivity in the mitogen-activated protein kinase cascade}
{\em Proc. Natl. Acad. Sci. USA}
{\bf 93}
\jyear{1996}
\pp{10078--10083} 

\bibitem{petri3}
R. K\"uffner,
R. Zimmer,
T. Lengauer,
\papertitle{Pathway analysis in metabolic databases via differential metabolic display (DMD)}
\journal{Bioinformatics}
\jvol{16}
\jyear{2000}
\pp{825-836}

\bibitem{lauffenburger}
D.A. Lauffenburger,
\papertitle{A computational study of feedback effects on signal dynamics in a mitogen-activated protein kinase (MAPK) pathway model}
{\em Biotechnol. Prog.}
{\bf 17}
\jyear{2001}
\pp{227--239}

\bibitem{kholodenko}
N.I. Markevich,
J.B. Hoek,
B.N. Kholodenko,
``Signaling switches and bistability arising from multisite phosphorilation in protein kinase cascades''
\emph{Journal of Cell Biology}, Vol. 164, N.3, pp.
353-359, 2004

\bibitem{petri5}
J.S. Oliveira,
C.G. Bailey,
J.B. Jones-Oliveira,
Dixon, D.A.,
Gull, D.W.,
Chandler, M.L.A.,
\papertitle{A computational model for the identification of biochemical pathways in the Krebs cycle}
\journal{J. Comput. Biol.}
\jvol{10}
\jyear{2003}
\pp{57-82}

\bibitem{petri6}
M. Peleg, M.,
I. Yeh,
R. Altman,
\papertitle{Modeling biological processes using workflow and Petri net models}
\journal{Bioinformatics }
\jvol{18}
\jyear{2002}
\pp{825-837}

\bibitem{peterson}
J.L. Peterson,
\book{Petri Net Theory and the Modeling of Systems}
Prentice Hall,
Lebanon, Indiana
1981.

\bibitem{ca_petri}
C.A. Petri,
\book{Kommunikation mit Automaten}
Ph.D. Thesis, University of Bonn,
1962.

\bibitem{reddy}
V.N. Reddy,
M.L. Mavrovouniotis,
M.N. Liebman,
\papertitle{Petri net representations in metabolic pathways.}
\journal{Proc. Int. Conf. Intell. Syst. Mol. Biol.}
\jvol{1}
\jyear{1993}
\pp{328-336}

\bibitem{hal1}
H.L. Smith,
\book{Monotone dynamical systems: An introduction to the theory of competitive and cooperative systems, Mathematical Surveys and Monographs, vol. 41}
(AMS, Providence, RI, 1995).

\bibitem{chemTAC}
\AUTHOR{E.D.}{Sontag}
\papertitle{Structure and stability of certain chemical networks and applications to the kinetic proofreading model of T-cell receptor signal transduction}
\journal{IEEE Trans.\ Autom.\ Control\/}
\jvol{46}
\jyear{2001}
\pp{1028--1047}
(Errata in {\it IEEE Trans.\ Autom.\ Control\/} {\bf 47}(2002): 705.)

\bibitem{mct}
\AUTHOR{E.D.}{Sontag}
\book{Mathematical Control Theory: Deterministic Finite Dimensional Systems, Second Edition}
Springer, New York
1998.

\bibitem{thieme}
H.R. Thieme,
\papertitle{Uniform persistence and permanence for non-autonomous semiflows in population biology}
\journal{Math. Biosci.}
\jvol{166}
\jyear{2000}
\pp{173-201}

\bibitem{widman}
C. Widmann,
G. Spencer,
M.B. Jarpe,
G.L. Johnson, G.L.,
\papertitle{Mitogen-activated protein kinase: Conservation of a three-kinase module from yeast to human}
{\em Physiol. Rev.}
{\bf 79}
\jyear{1999},
\pp{143--180}

\bibitem{schuster}
I. Zevedei-Oancea,
S. Schuster,
\papertitle{Topological analysis of metabolic networks based on Petri net theory}
\journal{In Silico Biol.}
\jvol{3}
\jyear{2003}
paper 0029.

\bibitem{zhou}
M. Zhou,
\book{Modeling, Simulation, and Control of Flexible Manufacturing Systems: A Petri Net Approach}
World Scientific Publishing,
Hong Kong,
1999.

\end{thebibliography}
\end{document}